# Lead, Follow, or Go Your Own Way:
## Empirical Evidence Against Leader-Follower Behavior in Electronic Markets

Karen Clay[*], Michael Smith[*], and Eric Wolff[**]

[*] *H. John Heinz III School of Public Policy and Management,*
*Carnegie Mellon University, Pittsburgh, Pennsylvania 15213*
*kclay@andrew.cmu.edu, mds@cmu.edu*

[**] *Graduate School of Industrial Administration,*
*Carnegie Mellon University, Pittsburgh, Pennsylvania 15213*
*wolff@andrew.cmu.edu*

This Version: September 24, 2001

*Abstract*

Low search costs in Internet markets can be used by consumers to find low prices, but can also be used by retailers to monitor competitors' prices. This price monitoring can lead to price matching, resulting in dampened price competition and higher prices in some cases. This paper analyzes price data for 316 bestselling, computer, and random book titles gathered from 32 retailers between August 1999 and January 2000. In contrast to previous studies we find no evidence of leader-follow behavior for the vast majority of retailers we study. Further, the few cases of leader-follow behavior we observe seem to be associated with managerial convenience as opposed to anti-competitive behavior. We offer a methodology that can be used by future academic researchers or government regulators to check for anti-competitive price matching behavior in future time periods or in additional product categories.

1. **Introduction**

Academic researchers have observed that electronic markets reduce buyer search costs, and that this reduction in search costs can lead to increased competition and lower prices among retailers (Bakos 1997). This observation led many commentators to predict that Bertrand competition would result, forcing prices to marginal cost and price dispersion to zero. Thus far, these predictions have not been realized. Studies of online book, compact disk, and software markets show that Internet prices are above cost and price dispersion in these markets is substantial (Bailey 1998; Brynjolfsson and Smith 2000; Smith, Bailey, and Brynjolfsson 2000; Clay, Krishnan and Wolff 2001).

However, electronic markets also reduce search costs for retailers when comparing prices with their competitors (Bakos 1998). Varian (2000) suggests that the ease with which retailers can observe their competitors' prices could lead to tacit collusion and prices above marginal cost. Such anti-competitive price matching was seen in early marketplaces established by airlines (Bloom, Milne, and Adler 1994; Foer 2001). In addition, recent studies have shown that well-known Internet retailers do tend to charge very similar prices for the same item (Clay, Krishnan and Wolff 2001, Smith 2001). The problem is determining whether parallel pricing by major retailers is a reflection of the competitive nature of the market or the market power of major retailers.

In the only study of parallel pricing thus far, Kauffman and Wood (2000) analyze pricing behavior by Internet retailers for books, CDs, and software products. They find evidence of leader-follower behavior in each product category, suggesting that firms may be raising prices above the competitive level. A significant limitation in their study is that it examines only products on industry bestseller lists. Since firms in these industries typically follow fixed pricing policies based on whether a product is listed in a bestseller list, it is impossible to determine whether the leader-follower behavior observed by the authors is due to endogenous leader-follower behavior or to reactions to exogenous changes in bestseller lists.

Our research uses a more extensive data set that includes both bestselling and non-bestselling titles to separate endogenous leader-follower behavior from reactions to exogenous changes in bestseller lists. We collected daily prices for 316 books from 32 online bookstores between August 1999 and August 2000. Our data include over 501,000 individual price observations and over 6,600 individual price changes. The sampled books included New York Times bestsellers, former New York Times bestsellers, computer bestsellers, former computer bestsellers, and books randomly selected from the *Books in Print* database. The stores included well-established Internet retailers such as Amazon, BarnesandNoble.com, and Borders.com as well as smaller Internet retailers such as Wordsworth and BCY Bookloft.

In contrast to previous studies, we find no evidence of leader-follow behavior for almost all the retailers we study. The one exception is Bookbuyers Outlet who appears to follow prices set by Amazon.com to the vast majority of titles they stock. However, this price matching behavior seems to be one of managerial convenience as opposed to anti-competitive.

The remainder of this paper proceeds as follows. Section 2 discusses the data gathered for this analysis. Section 3 presents our results, how these differ from previous work, and how they can be applied to future studies of leader-follower behavior in Internet markets. Section 4 concludes and discusses areas for future research.

## 2. Data

Our data include daily price observations for 316 books sold by 32 Internet retailers from August 1999 to January 2000. Books make a useful starting point for our analysis for several reasons. First, the Internet book market is relatively mature and has prominent retailers who make a natural focal point for leader-follower behavior. Second, books are a homogeneous product, which can be uniquely identified through an ISBN number.

Consistent with our desire to isolate endogenous leader-follower behavior from exogenous responses to changes in bestseller status, our book titles include 60 current and former New York Times bestsellers, 50 current and former computer bestsellers, and 206 books selected at random from *Books in Print*. The final dataset includes 501,193 individual price observations and 6,608 observed price changes over this period.

Data were collected using automated agents (spiders). Stores were included if they were covered in one of two major comparison-shopping engines – DealTime or PriceScan. These comparison-shopping engines do not cover the universe of all online stores, but the thirty-two stores in our sample cover the largest United States-based bookstores (i.e., Amazon, BarnesandNoble.com, Borders.com, Buy.com, and Booksamillion) and a representative sample of the smaller U.S.-based bookstores.[1]

The sample includes five categories of books: New York Times bestsellers, former New York Times bestsellers, computer bestsellers, former computer bestsellers, and a random sample of books listed in *Books in Print*. New York Times bestsellers were included, because they are widely carried, represent high aggregate sales, and are a focal point for discounts. We included all books appearing in the New York Times bestsellers lists for paperback fiction, paperback nonfiction, hardcover fiction and hardcover nonfiction for the week of August 8, 1999.[2] When New York Times bestsellers went off of the list, we continued to track them as former bestsellers.

---

[1] A complete list of the stores is provided in Table III. Data were also collected from some individual stores to confirm the accuracy of the information from DealTime and PriceScan.

Computer books were included, because they were one of the first categories of books sold on the Internet and remain an important category. Also, purchasers of computer books may be early adopters of comparison-shopping engines (Smith 2000). Although the New York Times does not maintain a bestseller list for computer books, Amazon does maintain such a list. We chose Amazon's bestseller list because of Amazon's high volume of book sales and the generalist (as opposed to specialist) orientation of the site.[3] Thus, the 50 books on the computer bestseller list are likely to be purchased by large numbers of consumers and offered in a large number of stores. Like the New York Times bestsellers, a panel was begun the week of August 8, 1999 and that panel was followed on an ongoing basis.

Random books were included to provide a baseline against which to compare the prices and price dispersion of bestsellers and former bestsellers and to understand pricing for the millions of books not covered by the bestseller lists. The random sample was created by generating random strings of letters of random length and then checking the result against the online *Books in Print* database until approximately 200 in-print titles were found.[4]

---

[2] The number is approximate, because there are often ties for the #15 spot.
[3] Use of any store's bestseller list raises unavoidable issues of endogeneity.
[4] Some, although technically in print, were not available in any bookstores. After eliminating these, the data set includes 181 random books.

Price changes are defined as a difference in listed price on an ISBN from one date to the next date with a listed price at the same store. Dates with missing prices are not considered as price changes. Furthermore, prices are only collected once per day. This methodology will undercount the true price changes to the extent that prices change several times during a single day. Many books that start on either the New York Times bestseller list or the Amazon list of computer bestsellers move off the bestseller lists during the sample period. During the sample period, many firms had a policy of decreasing the book prices as they move on the bestseller lists and increasing them as they move off the bestseller lists. However, stores do not appear to implement these changes on uniform dates relative to when the bestseller lists are formed and become public. In recognition of these lags in making price changes associated with movement on and off the lists, we consider price changes to New York Times bestsellers and Amazon computer bestsellers even after they move off their bestseller lists.

Summary statistics for the data set are presented in Table 1. The table records the distribution of price changes of random books, New York Times bestsellers, and computer bestsellers across the 32 sample stores. Across all three categories, Powell's made more changes than any other bookstore in the sample. The three largest players in the online market in terms of market share, Amazon, Barnesandnoble.com, and Borders, made far fewer price changes than Powell's did in each of these categories. For example, Powell's made 564 price changes to books in the random category while Amazon, Barnesandnoble.com, and Borders.com made 162, 96, and 135 changes, respectively. The variation in the number of price changes across stores suggests that changes in prices are not matched by all stores. Stores may be unaware of changes made by competitors in some cases, they make changes selectively according to unobserved criteria, or they may make changes independently of one another.

## 3. Results

To look for leader-follower behavior, we consider clusters of price changes related to single ISBN's. Hypothetically, a firm changes the price of a book. Other firms can respond immediately or wait to change their prices. As these firms change their prices, other firms, including the initial firm, may respond by changing prices. A single price change could result in a series of price changes. We use both 3-day and 7-day rules in forming price clusters. However, given that all retailers should be able to identify their competitors' price changes within hours using techniques similar to the ones we used to gather the data for this research, we find it more likely that price changes reflecting IT-enabled leader-follower behavior will fall within the 3-day rule.

Under the 3-day rule, we define a single price cluster as the sequence of price changes of a single ISBN that have three or fewer days between price changes. The initial price change comes four or more days after the last price change. The last price change in a cluster comes four or more days before the next price change. Table 2 shows the number of single ISBN price clusters in the sample using both 3-day (short window) and 7-day (long window) price clusters. For both the 3-day and 7-day clusters, the average length of the cluster in days is 2.7 and 6.9, respectively. Moreover, the average number of changes per cluster is 2.3 and 3.7 for 3-day and 7-day clusters respectively.

Table 3 provides additional detail on the store inclusion in 3-day clusters. For random books, computer books, and New York Times bestsellers, approximately 40% of the clusters have more than 5 stores. This result holds when looking at all changes or when restricting the focus to changes in the same direction (i.e. all price increases). While some clusters contain only one or two stores, price changes for ISBNs tend to take place at many online retailers.

One indication of leader-follower behavior is the distribution of stores that change prices on the first day of a price cluster[5]. Table 4 provides a distribution of initial price changes by store. This information is split further into single price changes (leaders with no followers) and the first of multiple price changes (leaders with followers). The ratio of single price changes to the first of multiple changes for individual stores provides a measure of influence by the firm. The lower this ratio, the more frequently the firm appears to be acting as a price leader. Using this metric, Amazon and Barnesandnoble.com are associated with price leadership. In contrast, varsitybooks.com and a1 books frequently change book prices, but these price changes are followed less frequently by competitors.

## 4.  Conclusions

One of the great promises of electronic commerce is that lower search costs would allow customers to easily locate the best price on a particular product resulting in "fierce price competition" (Kuttner 1998). However, more recently, researchers have observed that low search costs can also be used by retailers to follow their competitors price changes and that this may lead to increases in prices (Bakos 1998, Varian 2000).

This paper analyzes price data gathered from the Internet market for books from August 1999 to August 2000 to determine whether there is evidence to support leader-follower behavior in this

---

[5] The data do not permit examination of intraday price changes. We cannot tell from the data if one store changes its prices at 1:00pm and another changes at 2:00pm. Hence, we classify all stores that change prices on the first day of a cluster as having initiated the change.

Internet market. In contrast to previous studies of this phenomenon (Kauffman and Wood 2000), we find almost no evidence of leader-follower behavior among retailers. Furthermore, the one case of leader-follower behavior seems more consistent with managerial convenience as opposed to anti-competitive pricing. The likely difference between our study and previous work is that our study includes both bestselling and non-bestselling titles to separate endogenous leader-follower behavior from exogenous responses to changes in bestseller status.

However, more work is necessary by academic researchers and government regulators to confirm these results in additional product categories and in more recent data samples. We believe our methodology can be a useful starting point to these efforts.

# References

Bailey, Joseph P. 1998. Intermediation and electronic markets: Aggregation and pricing in Internet commerce. Ph.D., Technology, Management and Policy, Massachusetts Institute of Technology, Cambridge, MA.

Bakos, J. Yannis. 1997. Reducing buyer search costs: Implications for electronic marketplaces. *Management Science* **43**(12) 1613-1630.

Bakos, J. Yannis. 1998. The Emerging Role of Electronic Marketplaces on the Internet. *Communications of the ACM* **41**(8) 35-42.

Bloom, Paul N., Grorge R. Milne, Robert Adler. 1994. Avoiding misuse of new information technologies: Legal and societal considerations. *Journal of Marketing* **58**(1) 98-110.

Brynjolfsson, Erik, Michael Smith. 2000. Frictionless commerce? A comparison of Internet and conventional retailers. *Management Science* **46**(4) 563-585.

Clay, Karen, Ramayya Krishnan, Eric Wolff. 2001. Price strategies on the web: Evidence from the online book industry. *Journal of Industrial Economics* Forthcoming.

Foer, Albert A. 2001. E-Commerce Meets Antitrust: A Primer. *Journal of Public Policy and Marketing* **20**(1) 51-63.

Kauffman, Robert J., Charles A. Wood. 2000. Follow the Leader? Strategic Pricing in E-Commerce. *The Proceedings of the International Conference on Information Systems 2000*, Brisbane, Australia.

Kuttner, Robert. 1998. The Net: A Market Too Perfect for Profits? *Business Week*, May 11, p. 20.

Smith, Michael. 2000. Structure and Competition in Electronic Markets. Ph.D. Thesis, Management Science, Massachusetts Institute of Technology, Cambridge, MA.

Smith, Michael. 2001. The Law of One Price? Price Dispersion and Parallel Pricing in Internet Markets. Working Paper, Carnegie Mellon University, Pittsburgh, PA.

Smith, Michael, Joseph Bailey, Erik Brynjolfsson. 2000. Understanding digital markets. E. Brynjolfsson, B. Kahin, eds. *Understanding the Digital Economy* MIT Press, Cambridge, MA.

Varian, Hal R. 2000. Market Structure in the Network Age. E. Brynjolfsson, B. Kahin, eds. *Understanding the Digital Economy*, MIT Press, Cambridge, MA. Forthcoming.

**Table 1. Total Price Changes by Store**

| Store | Random | NYT | Computer |
|---|---:|---:|---:|
| 1000's of Discount Books | 0 | 0 | 0 |
| 1Bookstreet | 134 | 82 | 26 |
| A1Books | 357 | 171 | 123 |
| Allbooks4less.com | 4 | 0 | 0 |
| Alldirect.com | 28 | 0 | 2 |
| Alphacraze.com | 107 | 49 | 34 |
| Amazon | 162 | 122 | 64 |
| BarnesandNoble.com | 96 | 113 | 39 |
| BCY Bookloft | 237 | 0 | 112 |
| Bookbuyer's Outlet | 53 | 57 | 40 |
| Bookcloseouts | 0 | 1 | 0 |
| Bookpool.com | 17 | 0 | 38 |
| Books.com | 28 | 48 | 6 |
| Books4mom.com | 0 | 0 | 0 |
| Booksamillion.com | 260 | 124 | 44 |
| Booksnow.com | 45 | 14 | 20 |
| Borders.com | 135 | 94 | 80 |
| Brian's Books | 6 | 0 | 6 |
| Buy.com | 88 | 128 | 27 |
| Cherry Valley Books | 68 | 14 | 0 |
| Christianbook.com | 0 | 0 | 0 |
| Classbook.com | 90 | 17 | 34 |
| Cody's Books | 2 | 4 | 2 |
| Computerlibrary.com | 65 | 20 | 56 |
| Fatbrain | 202 | 75 | 108 |
| Hamiltonbook.com | 8 | 7 | 0 |
| Kingbooks.com | 38 | 4 | 6 |
| Page1book.com | 44 | 1 | 0 |
| Powells | 564 | 509 | 270 |
| Shopping.com | 321 | 181 | 102 |
| Varsitybooks.com | 174 | 44 | 84 |
| Wordsworth | 37 | 34 | 8 |
| *Total* | *3,370* | *1,913* | *1,331* |

Notes: A price change is defined as the difference in the reported price for a single ISBN by a single store from one price observation to the next.

**Table 2. Single ISBN Price Change Clusters**

|  | *3-day Cluster* | *7-day Cluster* |
|---|---|---|
| Total Number of Clusters | 2,858 | 1,781 |
| Avg. Length of Clusters (days) | 2.7 | 6.9 |
| Avg. Number of Price Changes Per Cluster | 2.3 | 3.7 |

Notes: N-day price clusters are defined for specific ISBN's. The initial price change for an n-day price cluster begins with the first price change occurring at least n-days after the last price change. Subsequent changes occurring within n-days of the last change are also included in the cluster. The end date of a cluster is the first date after the initial price change that has more than n days until the next price change.

# Table 3. Number of Stores in 3-Day Price Clusters[*]

| Random Books | | | |
|---|---|---|---|
| Number of stores in 3 day price cluster | Total changes | Changes up | Changes down |
| 1 store | 394 | 246 | 148 |
| 2 stores | 494 | 316 | 178 |
| 3 stores | 423 | 279 | 144 |
| 4 stores | 396 | 237 | 159 |
| 5 stores | 195 | 119 | 76 |
| + 5 stores | 1,583 | 924 | 659 |
| Computer Books | | | |
| Number of stores in 3 day price cluster | Total changes | Changes up | Changes down |
| 1 store | 120 | 78 | 42 |
| 2 stores | 162 | 101 | 61 |
| 3 stores | 189 | 115 | 74 |
| 4 stores | 136 | 85 | 51 |
| 5 stores | 140 | 86 | 54 |
| + 5 stores | 572 | 345 | 227 |
| Bestselling Books | | | |
| Number of stores in 3 day price cluster | Total changes | Changes up | Changes down |
| 1 store | 104 | 65 | 39 |
| 2 stores | 108 | 55 | 53 |
| 3 stores | 108 | 61 | 47 |
| 4 stores | 112 | 63 | 49 |
| 5 stores | 80 | 31 | 49 |
| + 5 stores | 1,316 | 821 | 495 |

Notes: Total changes is the total number of price changes listed by all stores in the price cluster. Changes up and changes down record the number of price changes within a 3-day price cluster in the same direction.

**Table 4. Distribution of Stores that Initiate Price Changes**

| Store | Single Price Changes | | 1st of Multiple Price Changes | |
|---|---|---|---|---|
| | 3 day change window | 7 day change window | 3 day change window | 7 day change window |
| 1000's of Discount Books | 0 | 0 | 0 | 0 |
| 1Bookstreet | 67 | 16 | 54 | 47 |
| A1Books | 265 | 174 | 176 | 156 |
| Allbooks4less.com | 0 | 0 | 3 | 1 |
| Alldirect.com | 8 | 5 | 3 | 3 |
| Alphacraze.com | 45 | 9 | 47 | 35 |
| Amazon | 21 | 9 | 147 | 87 |
| BarnesandNoble.com | 58 | 32 | 74 | 55 |
| BCY Bookloft | 55 | 31 | 115 | 101 |
| Bookbuyer's Outlet | 0 | 0 | 49 | 24 |
| Bookcloseouts | 0 | 0 | 0 | 0 |
| Bookpool.com | 17 | 5 | 22 | 16 |
| Books.com | 16 | 6 | 32 | 17 |
| Books4mom.com | 0 | 0 | 0 | 0 |
| Booksamillion.com | 53 | 11 | 124 | 105 |
| Booksnow.com | 1 | 0 | 25 | 20 |
| Borders.com | 64 | 26 | 78 | 66 |
| Brian's Books | 6 | 1 | 0 | 5 |
| Buy.com | 44 | 28 | 81 | 54 |
| Cherry Valley Books | 1 | 0 | 17 | 11 |
| Christianbook.com | 0 | 0 | 0 | 0 |
| Classbook.com | 11 | 9 | 69 | 57 |
| Cody's Books | 2 | 1 | 4 | 2 |
| Computerlibrary.com | 22 | 2 | 61 | 18 |
| Fatbrain | 81 | 24 | 111 | 85 |
| Hamiltonbook.com | 6 | 1 | 3 | 3 |
| Kingbooks.com | 18 | 4 | 13 | 7 |
| Page1book.com | 25 | 23 | 16 | 13 |
| Powells | 200 | 57 | 337 | 231 |
| Shopping.com | 128 | 38 | 194 | 123 |
| Varsitybooks.com | 136 | 98 | 70 | 53 |
| Wordsworth | 16 | 8 | 27 | 19 |

Notes: Single Price Changes records the number of price changes made by stores that occurred on the 1st day of a n-day cluster that were not followed by price changes by other stores. 1st of Multiple Price Changes records the number of times the price changes made by a store occurs on the first day of a n-day price cluster that was followed by price changes at other stores.

**Table 5A. Distribution of Price Changes Relative to Amazon's Price Changes: Random Books**

| | *Percentage of Amazon ISBN's carried by another store that changed prices* | | | | | | |
|---|---|---|---|---|---|---|---|
| | *Day of price change relative to price changes by Amazon* | | | | | | |
| *Store* | -3 | -2 | -1 | 0 | 1 | 2 | 3 |
| 1000's of Discount Books | 0% | 0% | 0% | 0% | 0% | 0% | 0% |
| 1Bookstreet | 0% | 0% | 0% | 2% | 0% | 8% | 21% |
| A1Books | 0% | 0% | 0% | 0% | 0% | 0% | 19% |
| Allbooks4less.com | 0% | 0% | 0% | 0% | 0% | 0% | 0% |
| Alldirect.com | 0% | 0% | 0% | 0% | 0% | 0% | 0% |
| Alphacraze.com | 0% | 0% | 0% | 1% | 13% | 0% | 4% |
| Amazon | 0% | 0% | 0% | 0% | 0% | 0% | 0% |
| BarnesandNoble.com | 6% | 0% | 15% | 3% | 9% | 5% | 7% |
| BCY Bookloft | 0% | 28% | 6% | 1% | 0% | 0% | 3% |
| Bookbuyer's Outlet | 0% | 0% | 0% | 54% | 36% | 8% | 0% |
| Bookcloseouts | 0% | 0% | 0% | 0% | 0% | 0% | 0% |
| Bookpool.com | 0% | 0% | 0% | 0% | 0% | 0% | 0% |
| Books.com | 0% | 0% | 0% | 0% | 0% | 6% | 7% |
| Books4mom.com | 0% | 0% | 0% | 0% | 0% | 0% | 0% |
| Booksamillion.com | 35% | 11% | 6% | 5% | 0% | 0% | 2% |
| Booksnow.com | 0% | 0% | 0% | 0% | 0% | 0% | 0% |
| Borders.com | 6% | 0% | 24% | 1% | 6% | 10% | 10% |
| Brian's Books | 0% | 0% | 0% | 0% | 0% | 0% | 0% |
| Buy.com | 13% | 5% | 12% | 1% | 0% | 6% | 9% |
| Cherry Valley Books | 0% | 0% | 0% | 0% | 0% | 0% | 0% |
| Christianbook.com | 0% | 0% | 0% | 0% | 0% | 0% | 0% |
| Classbook.com | 0% | 10% | 0% | 0% | 10% | 0% | 0% |
| Cody's Books | 0% | 0% | 6% | 0% | 0% | 0% | 0% |
| Computerlibrary.com | 0% | 0% | 0% | 1% | 0% | 0% | 0% |
| Fatbrain | 6% | 6% | 0% | 2% | 24% | 3% | 5% |
| Hamiltonbook.com | 0% | 0% | 0% | 0% | 0% | 0% | 0% |
| Kingbooks.com | 0% | 0% | 0% | 0% | 0% | 0% | 0% |
| Page1book.com | 0% | 0% | 0% | 0% | 0% | 0% | 0% |
| Powells | 11% | 6% | 6% | 2% | 19% | 5% | 2% |
| Shopping.com | 0% | 0% | 11% | 2% | 0% | 15% | 10% |
| Varsitybooks.com | 13% | 0% | 0% | 3% | 0% | 24% | 3% |
| Wordsworth | 0% | 0% | 0% | 0% | 0% | 0% | 0% |

Notes: Percentage of Amazon ISBN's carried by another store that changed is calculated as the number of price changes by another store for Amazon's ISBNs that changed price on date 0 divided by the total number of Amazon ISBNs that changed price on date 0 carried by another store.

**Table 5B. Distribution of Price Changes Relative to Amazon's Price Changes: New York Times Bestsellers**

| | *Percentage of Amazon ISBN's carried by another store that changed prices* | | | | | | |
|---|---|---|---|---|---|---|---|
| | *Day of price change relative to price changes by Amazon* | | | | | | |
| *Store* | *-3* | *-2* | *-1* | *0* | *1* | *2* | *3* |
| 1000's of Discount Books | 0% | 0% | 0% | 0% | 0% | 0% | 0% |
| 1Bookstreet | 0% | 4% | 14% | 5% | 29% | 26% | 15% |
| A1Books | 0% | 0% | 0% | 0% | 0% | 3% | 0% |
| Allbooks4less.com | 0% | 0% | 0% | 0% | 0% | 0% | 0% |
| Alldirect.com | 0% | 0% | 0% | 0% | 0% | 8% | 0% |
| Alphacraze.com | 0% | 0% | 0% | 1% | 0% | 0% | 2% |
| Amazon | 0% | 0% | 0% | 0% | 0% | 51% | 0% |
| BarnesandNoble.com | 0% | 5% | 20% | 20% | 36% | 0% | 5% |
| BCY Bookloft | 0% | 0% | 0% | 0% | 0% | 0% | 0% |
| Bookbuyer's Outlet | 0% | 0% | 0% | 58% | 0% | 0% | 0% |
| Bookcloseouts | 100% | 0% | 0% | 0% | 0% | 0% | 0% |
| Bookpool.com | 0% | 0% | 0% | 0% | 0% | 0% | 0% |
| Books.com | 8% | 7% | 8% | 0% | 0% | 9% | 4% |
| Books4mom.com | 0% | 0% | 0% | 0% | 0% | 0% | 0% |
| Booksamillion.com | 50% | 15% | 22% | 2% | 7% | 2% | 7% |
| Booksnow.com | 0% | 0% | 0% | 2% | 0% | 0% | 0% |
| Borders.com | 0% | 4% | 0% | 3% | 0% | 7% | 33% |
| Brian's Books | 0% | 0% | 0% | 0% | 0% | 0% | 0% |
| Buy.com | 0% | 5% | 5% | 3% | 31% | 26% | 15% |
| Cherry Valley Books | 0% | 0% | 0% | 0% | 0% | 0% | 0% |
| Christianbook.com | 0% | 0% | 0% | 0% | 0% | 0% | 0% |
| Classbook.com | 0% | 0% | 0% | 0% | 0% | 0% | 0% |
| Cody's Books | 0% | 5% | 0% | 0% | 0% | 0% | 0% |
| Computerlibrary.com | 11% | 0% | 0% | 0% | 0% | 0% | 0% |
| Fatbrain | 0% | 0% | 0% | 2% | 0% | 2% | 3% |
| Hamiltonbook.com | 0% | 0% | 0% | 0% | 0% | 0% | 0% |
| Kingbooks.com | 0% | 0% | 0% | 1% | 0% | 0% | 0% |
| Page1book.com | 0% | 0% | 0% | 0% | 0% | 0% | 0% |
| Powells | 5% | 23% | 5% | 4% | 7% | 3% | 8% |
| Shopping.com | 7% | 5% | 5% | 5% | 0% | 9% | 10% |
| Varsitybooks.com | 0% | 0% | 0% | 2% | 0% | 0% | 0% |
| Wordsworth | 0% | 0% | 0% | 0% | 0% | 0% | 0% |

Notes: Percentage of Amazon ISBN's carried by another store that changed is calculated as the number of price changes by another store for Amazon's ISBNs that changed price on date 0 divided by the total number of Amazon ISBNs that changed price on date 0 carried by another store.

**Table 5C. Distribution of Price Changes Relative to Amazon's Price Changes: Computer Bestsellers**

| | *Percentage of Amazon ISBN's carried by another store that changed prices* | | | | | | |
|---|---|---|---|---|---|---|---|
| | *Day of price change relative to price changes by Amazon* | | | | | | |
| *Store* | *-3* | *-2* | *-1* | *0* | *1* | *2* | *3* |
| 1000's of Discount Books | 0% | 0% | 0% | 0% | 0% | 0% | 0% |
| 1Bookstreet | 0% | 0% | 0% | 3% | 0% | 0% | 0% |
| A1Books | 20% | 0% | 0% | 5% | 0% | 0% | 50% |
| Allbooks4less.com | 0% | 0% | 0% | 0% | 0% | 0% | 0% |
| Alldirect.com | 0% | 0% | 0% | 0% | 0% | 0% | 0% |
| Alphacraze.com | 0% | 0% | 0% | 0% | 10% | 0% | 0% |
| Amazon | 0% | 0% | 0% | 0% | 0% | 0% | 0% |
| BarnesandNoble.com | 0% | 0% | 0% | 0% | 0% | 0% | 0% |
| BCY Bookloft | 0% | 20% | 0% | 11% | 0% | 0% | 0% |
| Bookbuyer's Outlet | 0% | 0% | 0% | 100% | 73% | 60% | 17% |
| Bookcloseouts | 0% | 0% | 0% | 0% | 0% | 0% | 0% |
| Bookpool.com | 0% | 0% | 0% | 0% | 0% | 0% | 0% |
| Books.com | 0% | 33% | 0% | 5% | 14% | 0% | 33% |
| Books4mom.com | 0% | 0% | 0% | 0% | 0% | 0% | 0% |
| Booksamillion.com | 0% | 0% | 0% | 0% | 8% | 0% | 0% |
| Booksnow.com | 0% | 0% | 0% | 0% | 0% | 0% | 0% |
| Borders.com | 0% | 0% | 0% | 2% | 0% | 0% | 0% |
| Brian's Books | 0% | 0% | 0% | 0% | 0% | 0% | 0% |
| Buy.com | 0% | 0% | 0% | 2% | 0% | 0% | 0% |
| Cherry Valley Books | 0% | 0% | 0% | 0% | 0% | 0% | 0% |
| Christianbook.com | 0% | 0% | 0% | 0% | 0% | 0% | 0% |
| Classbook.com | 0% | 0% | 0% | 0% | 0% | 0% | 0% |
| Cody's Books | 0% | 0% | 0% | 0% | 0% | 0% | 0% |
| Computerlibrary.com | 25% | 0% | 0% | 0% | 0% | 20% | 0% |
| Fatbrain | 0% | 17% | 33% | 0% | 0% | 0% | 0% |
| Hamiltonbook.com | 0% | 0% | 0% | 0% | 0% | 0% | 0% |
| Kingbooks.com | 0% | 0% | 0% | 0% | 0% | 0% | 0% |
| Page1book.com | 0% | 0% | 0% | 0% | 0% | 0% | 0% |
| Powells | 0% | 0% | 33% | 2% | 14% | 20% | 0% |
| Shopping.com | 0% | 20% | 0% | 8% | 0% | 20% | 0% |
| Varsitybooks.com | 0% | 0% | 0% | 0% | 0% | 0% | 0% |
| Wordsworth | 0% | 100% | 0% | 3% | 0% | 0% | 0% |

Notes: Percentage of Amazon ISBN's carried by another store that changed is calculated as the number of price changes by another store for Amazon's ISBNs that changed price on date 0 divided by the total number of Amazon ISBNs that changed price on date 0 carried by another store.